\documentclass[referee,useAMS,usenatbib,usegraphicx]{mn2e}
\usepackage{amssymb}
\usepackage{graphicx}

\title[Imaging polarimetry of comets C/2013 V1 (Boattini) and 290P/Jager]
{Imaging polarimetry of comets C/2013 V1 (Boattini) and 290P/Jager before and after perihelion}
\author[P. Deb Roy, P. Halder,  H. S. Das and  B. J. Medhi ]
{P. Deb Roy$^{1}$
, P. Halder$^{1}$
, H. S. Das$^{1}$\thanks{E-mail: hsdas@iucaa.ernet.in (HSD)}
and B. J. Medhi$^{2}$
\\
$^{1}$ Department of Physics, Assam University, Silchar 788011, India.\\
$^{2}$ Aryabhatta Research Institute of Observational Sciences, Manora Peak, Nainital 263129, India\\
}

\begin{document}

\date{Accepted 2015 March 27. Received 2015 January 9; in original form 2014 October 19}

\pagerange{\pageref{firstpage}--\pageref{lastpage}} \pubyear{2015}

\maketitle

\label{firstpage}

\begin{abstract}
We report the results obtained from the optical polarimetric study of the light scattered by comets C/2013 V1 (Boattini) and 290P/Jager at lower phase angles. The polarimetric observations of two comets have been performed with the 1.04-m Sampurnanand telescope of Aryabhatta Research Institute of observational sciencES near Nainital in India on 2013 December 4 and 5 and on 2014 April 24 using R photometric band ($\lambda$ =  630 nm, $\Delta$$\lambda$ =120nm). We covered observations in both the pre and post perihelion passage of comets C/2013 V1 (Boattini) and 290P/Jager at two phase angles $\sim$ 13$^\circ$ and 27$^\circ$. The degree of polarization changes from ($-1.4$$\pm 0.3$)per cent to (+2.8$\pm 0.5$)per cent for comet C/2013 V1 (Boattini) and ($-1.6$$\pm 0.5$)per cent to (+2.5$\pm 0.5$)per cent for comet 290P/Jager at phase angles $\sim$ 13$^\circ$ and 27$^\circ$ respectively. The change in the physical properties of cometary dust is being well studied from the polarization maps obtained for both the period of observations. It is found that the aperture polarization values are comparable to those of other comets. The variation in the brightness profile of both the comets from the standard canonical nature is also being observed in both the solar and anti-solar direction during this phase which suggests the various physical evolution influencing the cometary comae.
\end{abstract}


\section{Introduction}
Comets are small icy Solar system body spend most of their lifetime far away from the Sun, hence the material on their subsurface is assumed to be primordial. Linear polarization measurements of the scattered solar radiation is a primary tool to study atmosphere less celestial bodies and comets. The main objective behind the polarimetric study of comets C/2013 V1  (Boattini) and 290P/Jager is to understand the physical properties of dust grains present in the cometary coma.

Michael Jager discovered the Jupiter Family Comet (JFC) 290P/Jager with a 0.25-m f/2.8 Schmidt camera on 1988 October 25. Recently, non-periodic comet C/2013 V1  (Boattini) was discovered by A. Boattini  with the Catalina Sky Survey's 0.68-m Schmidt telescope on 2013 November 6. The orbital elements of the comet C/2013 V1 (Boattini) indicate a perihelion passage on 2014 April 21 at a distance of 1.66 au and that of comet 290P/Jager suggests its perihelion date to be 2014 March 12 at a distance of 2.16 au respectively from the Sun. The orbital plane of comet C/2013 V1  (Boattini) and  290P/Jager inclined at an angle of  65$^\circ$.3 and 19$^\circ$.1 respectively with the ecliptic plane.

An extensive study of cometary dust grains has been done by various researchers, but the ambiguity over their exact nature and composition still prevails. Hence to clear this enigma several investigators prefer linear polarization measurements and theoretical modelling in order to reveal the true characteristics of cometary dust grains (Das, Sen \& Kaul 2004, Jockers et al. 2005, Das \& Sen 2006, 2011; Kimura, Kolokolova \& Mann 2006; Levasseur-Regourd et al. 2007; Das et al. 2008a; Das, Das \& Sen 2008b, 2010, 2011; Lasue et al. 2009; Hadamcik et al 2010, Kiselev et al. 2013, etc.).

In the past, we conducted CCD polarimetric observations of comet C/2009 P1 (Garradd) (Das et al. 2013) and comet C/2012 L2 (LINEAR) (Deb Roy, Das \& Medhi 2015) at lower phase angles to study the dust evolution in the extended coma around the nucleus. Both the comets showed a noticeable variation in the intensity and polarization profile in all considered direction with a prominent jet ejection in the solar and anti-solar direction respectively.  However, a negative polarization of ($-1$$\pm 0.7$)\%  was observed  at phase angle of about 21$^\circ$.6 for comet C/2009 P1 (Garradd). Recent observation of comet C/2012 L2 (LINEAR) showed a change in the direction of jet extension in both the intensity and polarization map between two consecutive nights of observation which might be due to the rotation of the nucleus with a whole coma polarization of ($+2.8$$\pm 0.3$)\% at 31$^\circ$.1 phase angle.

The cometary dust grains present in the coma of a comet undergoes continuous evolution with distance from the nucleus during the entire period of cometary activity. Polarimetric studies of comets reveal the evolutionary characteristics of comet dust grains in the coma. The observations of both the comets were conducted at lower phase angles. Actually, the polarimetric data of comets at large phase angles are numerous, but there is a lack of observations made at small phase angles. Observation of comet at lower phase angle is a typical task as comet is expected to be far off in the sky and appears as faint object at such phase angle. But the polarimetric study at lower phase angle helps to explore the polarimetric behaviour of comets especially in the negative branch of polarization--phase curve.

In the present study, we report observations in both the pre and post perihelion passage of comets C/2013 V1  (Boattini) and 290P/Jager; and the results obtained from the polarimetric analysis of both comets to explore the observed variation in both the brightness and polarization distribution across the coma during the phase 2013 December and 2014 April when the phase angle varies from 13$^\circ$ to 27$^\circ$.


\section{OBSERVATIONS AND REDUCTION OF THE POLARIMETRIC DATA}

The  Aryabhatta Research Institute of observational sciencES (ARIES) near Nainital in India is located at  79$^\circ$27$^\prime$ E,  29$^\circ$22$^\prime$  N at an altitude of 1951 m. The polarimetric observations were performed with the 1.04-m Sampurnanand telescope of ARIES  on 2013 December 4 and 5 and on 2014 April 24. The observations were carried out covering both pre and post perihelion passage of comets C/2013 V1  (Boattini) and 290P/Jager. The observational data are delimitated in [Table 1]. Both the comets were located at almost same phase angles in the months of 2013 December and 2014 April. The 1.04-m ARIES Sampurnanand Telescope (AST) is equipped with ARIES Imaging Polarimeter (AIMPOL) as a focal plane instrument. AIMPOL makes  use of a  Wollaston prism and a half-wave plate (HWP). The incident beam gets split by Wollaston prism into an ordinary and an extraordinary component. A CCD camera of 1024 x 1024 pixels is used to record the images. Broad-band red filter ($\lambda$ = 630  nm, $\Delta$$\lambda$=120nm) is used during all the phase of observation which is more frequent to observe the fainter objects. The resolution of each pixel of the CCD is 1.73 arcsec with an effective field of view of about 8 arcmin  diameter on the sky. The gain and read-out noise of the CCD used for observation are 11.98  e$^-$/ADU and 7.0 e$^-$ respectively. The effective seeing radius is $\sim$ 3 arc sec during December and $\sim$ 4 arcsec in April month of observation. The detailed description of the instrument is given in Medhi et al. (2010), Das et al. (2013) and references there in.

The CCD recorded images were processed with standard scientific procedures which includes bias subtraction, flat-field correction and the determination of instrumental polarization parameters. All the  bias subtracted images used for photopolarimetric analysis are divided by the master flat-field obtained from all the sky flats observed at twilight. Sky background is a serious task in comet analysis since the dusty coma of comet extends over a large distance. For proper subtraction of sky counts we subtracted a constant value that was estimated at the edge of each frame far from the comet photocentre free from faint stars where the contribution from cometary coma was negligible. The photocentre of each polarized image of both the comets was found with a precision of 0.1 pixel. The alignment of subimages with different angles of rotation of HWP during a particular night of observation have been done and then co-added which enhanced the sensitivity to coma detection and finally constructed the intensity and polarization map with the properly aligned images.

Four standard stars with high and three stars with low degree of polarization were observed to determine the instrumental polarization and the zero-point of the position angle of polarization plane. The standard stars for null  polarization and  zero-point of the polarization position angle were taken from Schmidt et al. (1992), Turnshek et al. (1990) and HPOL\footnote[1]{http://www.sal.wisc.edu/HPOL/tgts/HD251204.html}, respectively. After the instrumental polarization correction the computed polarization degree of the standard stars differs from the literature values by 0.1\% and the offset in position angle of plane of polarization ($\theta_0$ = $\theta$ - $\theta$$_{obs}$) by $-7^\circ$. The polarization details of the standard stars are depicted in Table 2.

The linear polarization ($p$) and position angle of the polarization vector ($\theta$) interms of the normalized stoke's parameters $q = Q/I$ and $u = U/I$ is given by

\begin{equation}
p = \sqrt{(q^2 + u^2)} ~~~~ \textrm{and} ~~~~ \theta = 0.5~\textrm{tan}^{-1}(q/u)
\end{equation}

Basically the position angle of the plane of polarization $({\theta}_{r})$ with respect to the normal of the scattering plane determines the sign of the polarization. The value ${p}_{r}$ and ${ \theta}_{r}$ referred to the scattering plane are connected with the quantities ${p}_{obs}$ and ${\theta}_{obs}$ through the relation

 \begin{equation}
P_r = P_{obs}. \cos 2\theta_r, ~~~~\theta_r = \theta_{obs} - (\phi \pm 90^\circ)
\end{equation}

    \ where $\phi$ is the position angle of the scattering plane and the sign in the bracket is chosen to ensure the condition 0$\leq$ $\phi$ $\pm$ 90$^o$ $\leq$ 180$^o$. During the observation in December the average value of ${ \theta}_{r}$ is computed to be 79$^\circ$ for comet C/2013 V1 (Boattini) and 89$^\circ$ to that of comet 290P/Jager. So the polarization values will be negative during this phase which is quite expected at the phase angle close to 13$^\circ$.


\begin{table*}
\caption{Observation log. comet, Observation date,  geocentric distance ($\triangle$),heliocentric distance (r), phase angle ($\alpha$), position angle of extended Sun -- comet radius
vector ($\phi$), projected diameter for 1 pixel (D) during the observations. }
\begin{center}
\begin{tabular}{|c|c|c|c|c|c|c|c|c|c|}
\hline
    Comet & Observation date & $\triangle$ (au)&  r (au) & $\alpha$ ($^\circ$)  & $\phi$ ($^\circ$)& D  \\
               &         &       &                  &        &                      & (km  px$^{-1}$) \\
 \hline
C/2013 V1 (Boattini)   &  2013 December 4 & 1.54 & 2.42 & 12.8 & 15.9 & 1927\\
                   &   2013 December 5 & 1.53 & 2.41 & 12.9 & 18.6 & 1918\\
                   &  2014 April 24 & 2.18 & 1.66 & 26.1 & 40.3 & 2740\\
                   \hline

290P/Jager        &  2013 December 4 & 1.45 & 2.34 & 13.1 & 244.5 & 1818\\
                  &  2013 December 5 & 1.44 & 2.34 & 12.8 & 243.4 & 1807\\
                  & 2014 April 24 & 2.15 & 2.19 & 26.9 & 100.6 & 2694\\
 \hline
\end{tabular}
\end{center}
\end{table*}


\begin{table*}
\caption{Results of standard polarized and unpolarized stars at R-filter. p and $\theta$ from the literature
Serkowski (1974); Turnshek et al. (1990) and HPOL. p$_{obs}$ and $\theta$$_{obs}$ from observations. Offset angle is calculated using the relation: $\theta_0$ = ($\theta - \theta_{obs}$).}
\begin{center}
\begin{tabular}{|c|c|c|c|c|c|c|c|c|c|}
\hline
    & Obs. date & Star  & p (\%) & $\theta$ ($^\circ$)   & p$_{obs} (\%)$ & $\theta$$_{obs}$ ($^\circ$) & $\theta _0$ \\
 \hline
   &  2013 December 4 & HD25543   & 4.73$\pm 0.05$  & 133.6     & 4.85$\pm 0.05$ & 141$\pm 2$  & --7.4$^\circ$ \\
      &                 & HD236633  & 5.37$\pm 0.03$  & 93.04     & 5.32$\pm 0.07$ & 100$\pm$2 & --6.9$^\circ$\\
      &                 &           &               &              &       &               & \\
       &   & HD21447  & 0.017$\pm 0.03$  & 28.3     & 0.13$\pm 0.04$ & 23$\pm 4$  & 5.3$^\circ$ \\
      &                 & HD14069  & 0.02$\pm 0.02$  & 156.6     & 0.06$\pm 0.03$ & 139$\pm$8 & 17.6$^\circ$\\
      &                 &           &               &              &       &               & \\
      &  2013 December 5 & HD25543 & 4.73$\pm 0.05$ & 133.6   & 4.83$\pm 0.05 $ & 140.8$\pm2$&  --7.2$^\circ$ \\

      &                 & HD236633   & 5.37$\pm 0.03$  & 93.04     & 5.34$\pm 0.07$ & 100$\pm$2 & --6.9$^\circ$\\
      &                 &           &               &              &       &               & \\
       &   & HD21447   & 0.017$\pm 0.03$  & 28.3     & 0.12$\pm 0.03$ & 23.1$\pm 4$  & 5.2$^\circ$ \\
      &                 & HD14069  & 0.02$\pm 0.02$  & 156.6     & 0.06$\pm 0.03$ & 138.7$\pm$8 & 17.9$^\circ$\\
 &                 &           &               &              &       &               & \\
      &  2014 April 24 & HD161056 & 4.05$\pm 0.05$ & 59.1   & 4.0$\pm0.01 $ & 66.3$\pm0.3$&  --7.2$^\circ$ \\

      &                 & GD319   & 0.09$\pm 0.09$  & 140.0    & 0.15$\pm0.08$ & 142.4$\pm$9 & --2.4$^\circ$\\

 \hline
\end{tabular}
\end{center}
\end{table*}



\begin{figure*}
\hspace{-1.5cm}
\includegraphics[width=150mm]{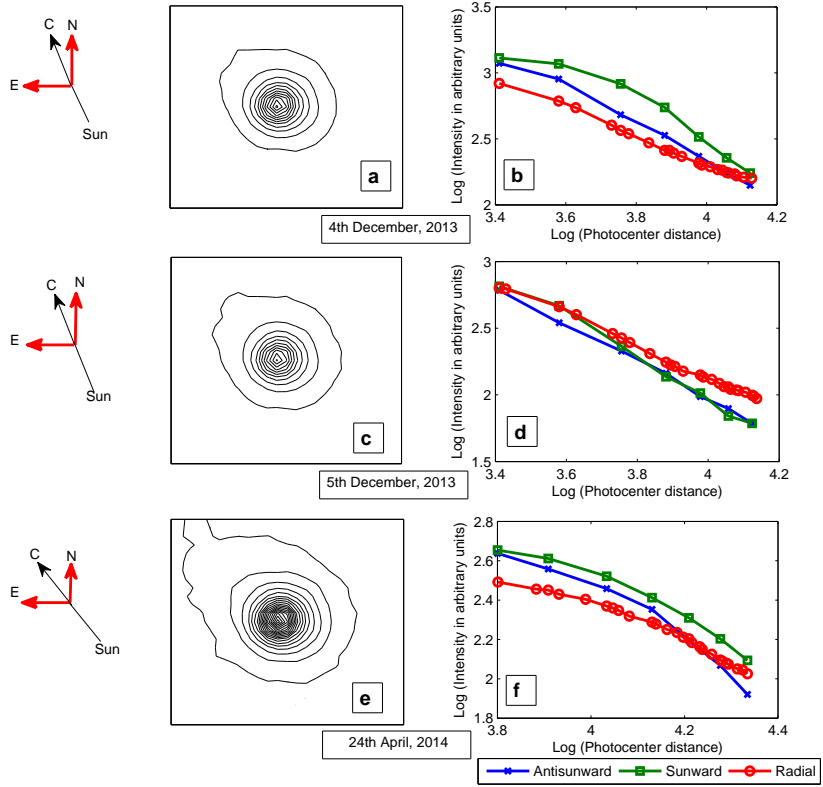}
 \vspace{0.5cm}
\caption{(a), (c) and (e): intensity map of comet C/2013 V1 (Boattini) with contours on 2013 December 4, 5 and 2014 April 24; (b), (d) and (f): the cuts through the coma sun-ward, anti-sunward and radial profile for the corresponding periods of observations. The first column of the plot shows the position angle of the Sun-comet radius vector which is 15$^\circ$.9, 18$^\circ$.6 and 40$^\circ$.3 respectively. The field of view is about 45,000 $\times$ 45,000 km
}
\end{figure*}


\begin{figure*}
\hspace{-1.5cm}
\includegraphics[width=150mm]{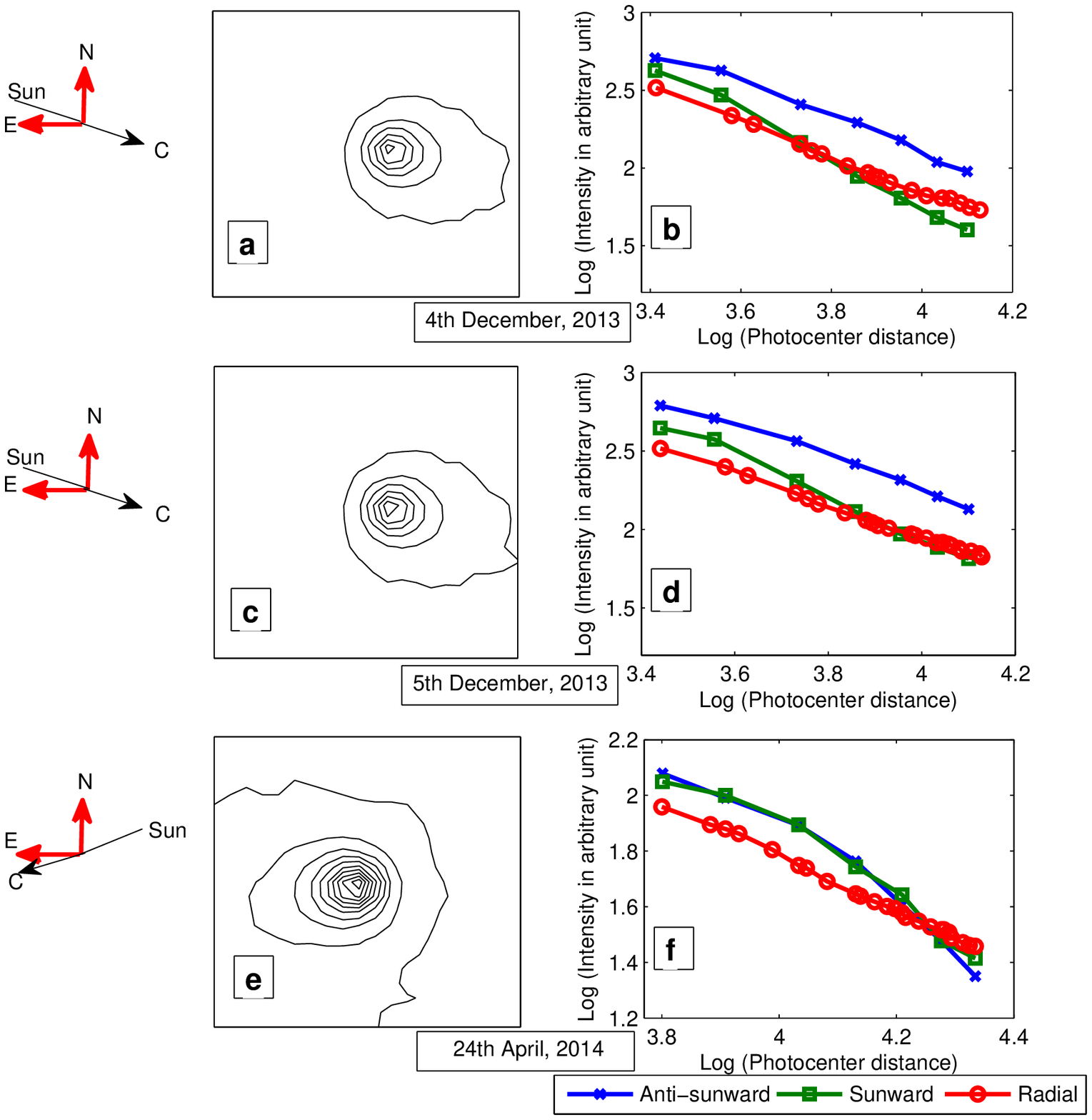}
 \vspace{0.5cm}
\caption{(a), (c) and (e): intensity map of comet 290P/Jager with contours on 2013 December 4, 5 and 2014 April 24; (b), (d) and (f): the cuts through the coma sun-ward, anti-sunward and radial profile for the corresponding periods of observations. The first column of the plot shows the position angle of the Sun-comet radius vector which is 244$^\circ$.5, 243$^\circ$.4 and 100$^\circ$.6 respectively. The field of view is about 45,000 $\times$ 45,000 km}
\end{figure*}


\section{RESULTS AND ANALYSIS}

   \subsection{Brightness Profile}
The surface brightness profile of the cometary coma falls with the photocentric distance. The azimuthally integrated profile along with the profiles in other specified direction explores the various physical evolution influencing the cometary coma. [Figs 1 \& 2] illustrate the brightness profile for comets C/2013 V1 (Boattini) and 290P/Jager during December and April respectively.

The seeing radius for both the comets in the month of 2013 December and 2014 April is 2700 km and 6700 km, respectively. The brightness profile of non-periodic comet C/2013 V1 (Boattini) shows almost opposite trend to that of JFC 290P/Jager in all phase of observation. The average radial profile of comets C/2013 V1 (Boattini) and 290P/Jager falls off as a function of photocentric distance with a slope of ($-1.0\pm0.08$), a typical feature of the expanding dusty coma around the nucleus of the comet. The slopes for sunward and anti-sunward directions in the month of December and April are shown in [Table-3]. The fall in intensity is higher in the anti-sunward direction for comet C/2013 V1 (Boattini) on 2013 December 4, when distance to the photocentre ($d$) is $\lesssim$ 8000 km. This fall is almost similar for both the directions when $d$ is more than 8000 km. On December 5, a higher fall in intensity in sunward direction is noticed for the same comet when $d$ is between 6000 and 8000 km.  Comet 290P/Jager shows almost reverse trend where the fall in intensity is found to be higher in the sunward direction during the observation on December 4 and 5. An identical profile is being observed in April phase of observation for both the comets where a highly steep fall in the intensity is detected in the anti-sunward direction as compared to the sunward direction mainly at a distance between 15,000 km and 21,000 km. The deviation of the brightness profile in the sunward and anti-sunward direction from the standard canonical nature confirms the time variable dust outflow. Different evolutionary processes such as solar radiation pressure, sublimation of icy grains, modulation of dust production rate and the change in the optical properties of dust grains could be the main reason behind the steepen or flattening out the brightness profile.


\begin{table*}
\caption{Intensity profile variation along sunward and anti-sunward direction on different date of observations throughout the coma. Here, $d$ is the distance to the photometric centre in kilometre.}.
\begin{center}
\hspace*{-1cm}
\begin{tabular}{c c c c c c c c c}
\hline \hline
$d$  & $\rightarrow$         & \    & \ 3000        & \ 6000
         &   \  8000 &   \ 12000  & \ 15000 &   \ 21000      \\

     (in km)   & Comet    &        &           & ~                  & ~
         & ~             & ~         & ~      \\
\hline
& \underline{\textbf{Sunward}}\\
 December 4, 2013 & 290P/Jager   &  & $-$1.10$\pm$0.12 & $-$1.35$\pm$0.11 & $-$1.70$\pm$0.18  & $-$1.51$\pm$0.19 & -- & -- \\
  & C/2013 V1 (Boattini) & & $-$0.60$\pm$0.10 & $-$0.75$\pm$0.11 & $-$0.98$\pm$0.13  & $-$1.75$\pm$0.20 & -- & --   \\

& \underline{\textbf{Antisunward}}\\
 December 4, 2013 & 290P/Jager  & & $-$0.75$\pm$0.11 & $-$0.88$\pm$0.11 & $-$1.10$\pm$0.13  & $-$1.28$\pm$0.12 & -- & -- \\
  & C/2013 V1 (Boattini) & & $-$0.72$\pm$0.09 & $-$0.95$\pm$0.09 & $-$1.34$\pm$0.13  & $-$1.78$\pm$0.16 & -- & --  \\

  \hline

 &  \underline{\textbf{Sunward}}\\
 December 5, 2013 & 290P/Jager  & & $-$1.05$\pm$0.10 & $-$1.20$\pm$0.11 & $-$1.50$\pm$0.11  & $-$1.36$\pm$0.13 & -- & -- \\
  & C/2013 V1 (Boattini) & & $-$0.85$\pm$0.10 & $-$1.24$\pm$0.10 & $-$1.52$\pm$0.12  & $-$1.40$\pm$0.13 & -- & --  \\

& \underline{\textbf{Antisunward}}\\
 December 5, 2013 & 290P/Jager  &  & $-$0.70$\pm$0.09 & $-$0.85$\pm$0.10 & $-$1.05$\pm$0.10  & $-$1.20$\pm$0.12 & -- & -- \\
  & C/2013 V1 (Boattini) & & $-$1.35$\pm$0.10 & $-$1.32$\pm$0.12 & $-$1.30$\pm$0.10  & $-$1.50$\pm$0.12 & -- & --    \\

  \hline

  & \underline{\textbf{Sunward}}\\
 April 24, 2014 & 290P/Jager & & -- & -- & $-$0.80$\pm$0.10
  & $-$1.02$\pm$0.10 & $-$1.30$\pm$0.15 & $-$1.58$\pm$0.15 \\
  & C/2013 V1 (Boattini) & & -- &-- & $-$0.70$\pm$0.09 & $-$0.85$\pm$0.10 & $-$1.22$\pm$0.13  & $-1$.60$\pm$0.18    \\

& \underline{\textbf{Antisunward}}\\
 April 24, 2014 & 290P/Jager &  & -- & -- & $-$0.89$\pm$0.10
  & $-$1.12$\pm$0.11 & $-$1.36$\pm$0.16 & $-$1.92$\pm$0.23 \\
  & C/2013 V1 (Boattini) & & -- & -- & $-$0.85$\pm$0.12
  & $-$0.94$\pm$0.14 & $-$1.28$\pm$0.19 & $-$1.84$\pm$0.22 \\

  \hline
\end{tabular}\label{tab2}
\end{center}
\end{table*}


\subsection{Polarization profile}

\subsubsection{Aperture Polarization}

The aperture polarization values over the whole coma of comets C/2013 V1 (Boattini) and 290P/Jager during all the cycles of observation are depicted in Table 4. The observed polarization for comet C/2013 V1 (Boattini) is found to be ($-1.4 \pm 0.3$)\% at 12$^\circ$.8 phase angle whereas 290P/Jager shows ($-1.6 \pm 0.3$)\% polarization at phase angle 13$^\circ$ on the first night. The two comets show almost identical polarization on the second night of observation during December. Both the comet shows high positive polarization at phase angle of about 27$^\circ$ in the April cycle of observation.


\begin{table*}
\caption{Polarization values for different apertures (diameters) and the position angle ($\theta_{obs}$) of polarization vector at R filter.}
\begin{tabular}{|c@{\hskip 0.05cm}|c@{\hskip 0.05cm}|c@{\hskip 0.05cm}|c@{\hskip 0.05cm}|c@{\hskip 0.05cm}|c@{\hskip 0.05cm}|c@{\hskip 0.05cm}|c@{\hskip 0.05cm}|c@{\hskip 0.05cm}|c@{\hskip 0.05cm}|}
\hline
   Diameter (in km) & $\rightarrow$  & 6000 & 12000  & 15000 & 21000 & 25000 & 30000 & 40000 & $\theta_{obs}$  \\
   Observation date & Comet  & &   &  &  &  &  & &(in degrees)\\
 \hline
 2013 December 4 & 290P/Jager  & $-$1.4$\pm$0.4 & $-$1.5$\pm$0.4 & $-$1.5$\pm$04. & $-$1.6$\pm$0.5   &--  &--  & -- & 65.9$\pm$10.1\\
&  C/2013 V1 (Boattini)  & $-$1.5$\pm$0.3 & $-$1.5$\pm$0.3 & $-$1.4$\pm$0.3 & $-$1.4$\pm$0.4   &-- &--  &--  & 27.3$\pm$7.0\\
 \hline
 2013 December 5 & 290P/Jager  & $-$1.3$\pm$0.3 & $-$1.3$\pm$0.3 & $-$1.4$\pm$0.3 & $-$1.4$\pm$0.3   &-- &-- &-- & 70.8$\pm$4.6\\
 & C/2013 V1 (Boattini)  & $-$1.4$\pm$0.3 & $-$1.4$\pm$0.3 & $-$1.3$\pm$0.3 & $-$1.4$\pm$0.4   &--  &--  &-- &22.8$\pm$5.5\\
 \hline
2014 April 24 & 290P/Jager &-- &--  & 2.4$\pm$0.5 & 2.4$\pm$0.5 & 2.5$\pm$0.5 & 2.4$\pm$0.6   & 2.2$\pm$0.6 & 147.2$\pm$6.1\\
  & C/2013 V1 (Boattini)  &--  &--  & 2.8$\pm$0.5 & 2.8$\pm$0.5   & 2.7$\pm$0.6 &2.8$\pm$0.6 & 2.7$\pm$0.7 & 173.3$\pm$7.2 \\
 \hline
\end{tabular}
\end{table*}

\subsubsection{Polarization map}

The polarization map is being built up with the properly aligned polarized components corresponds to four angles of rotation of HWP. The map derived by CCD imaging polarimetric technique offers a spatial resolution to the inner coma of the comets. [Fig 3 \& 4] demonstrate the polarization maps for comets C/2013 V1 (Boattini) and 290P/Jager on December 4 and April 24 respectively.

The comet C/2013 V1 (Boattini) shows high negative polarization of about $-3$\% at the centre as compared to the whole coma polarization of about ($-1.4 \pm 0.3$)\%. The polarization is highly negative for comet 290P/Jager of about $-4.5$\% at its centre while the circumnucleus halo possesses $-3 to $-2 per cent polarization which is found to extend in the antisolar direction. But positive polarization of the order of 0 - 1.5\% is found to be dominated in the outer comae of the comets suggests the presence of smaller grains in the extended coma of the comet as compared to the existence of more submicronic absorbing grains of complex texture in the circumnucleus region of the comet which is responsible for the observed negative polarization. The negative polarization region is found to be extended over a distance of $\sim$ 10,000 km from the photocentre in the anti solar direction for comet 290P/Jager where any such special feature is quite absent in comet C/2013 V1 (Boattini) during the observation in December.

In April, the polarization map shows a completely opposite trend where both the comet possesses a high positive polarization near the photocentre of about 3 per cent while it is found to vary between 2 and 1 per cent in the outer coma of the comet  as compared to the whole coma polarization of about (2.8 $\pm 0.5$)per cent for comet C/2013 V1 (Boattini) and (2.5 $\pm 0.5$)per cent for comet 290P/Jager. The observed variation in polarization over the whole coma for both the comets is due to the change in the physical properties of the dust grains in different regions of the coma.

To summarize, the polarization is highly negative in the circumnucleus halo of comet 290P/Jager as compared to that of comet C/2013 V1 (Boattini)  during December observation while a high positive polarization is being observed at the near nucleus region of both the comets in April observation.


\begin{figure}
\vspace{1cm}
\includegraphics[width=90mm]{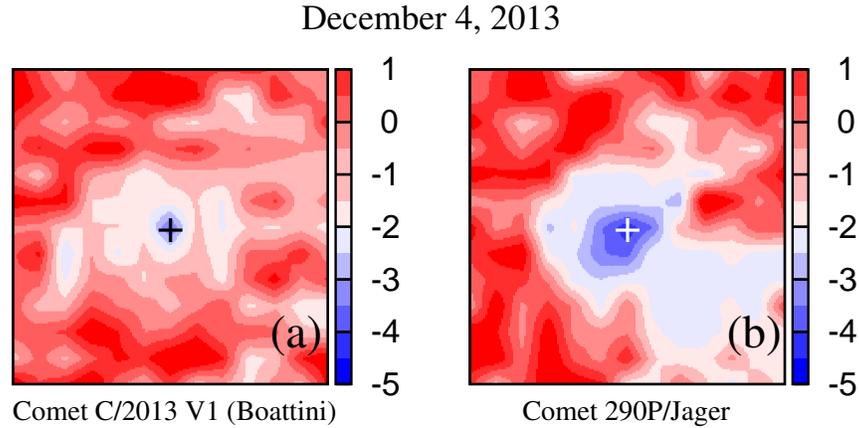}
\caption{ The polarization maps of (a) comet C/2013 V1 (Boattini) at 12$^\circ$.8 and (b) comet 290P/Jager at 13$^\circ$.1 phase angle.  The `+' mark denotes the photocentre of  the comet. The field of view is about 23,000 $\times$ 23,000 km.}
\end{figure}



\begin{figure}
\vspace{1cm}
\includegraphics[width=90mm]{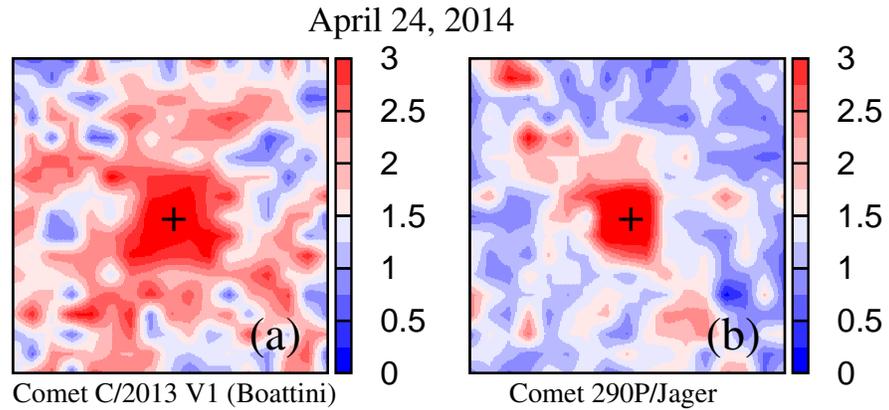}
\caption{ The polarization maps of (a) comet C/2013 V1 (Boattini) at 26$^\circ$.1 and (b) comet 290P/Jager at 26$^\circ$.9 phase angle.  The `+' mark denotes the photocentre of  the comet. The field of view is about 45,000 $\times$ 45,000 km.  }
\end{figure}


\subsubsection{Polarimetric phase curves}

In [Fig. 5] we combine the polarimetric phase curve for various comets with our observed data for comets C/2013 V1 (Boattini) and 290P/Jager at similar phase angle and we found good agreement of our computed polarization value with those reported by other investigators for different comets at such phase angles. Typically phase angle dependence of polarization contains a negative branch at lower phase angles (below 20$^\circ$) and a positive branch at higher phase angles. We found negative polarization for both the comets in the month of December at phase angle near 13$^\circ$ before their perihelion passage.

\begin{figure*}
\vspace{0.5cm}
\hspace{0.8cm}
\includegraphics[width=150mm]{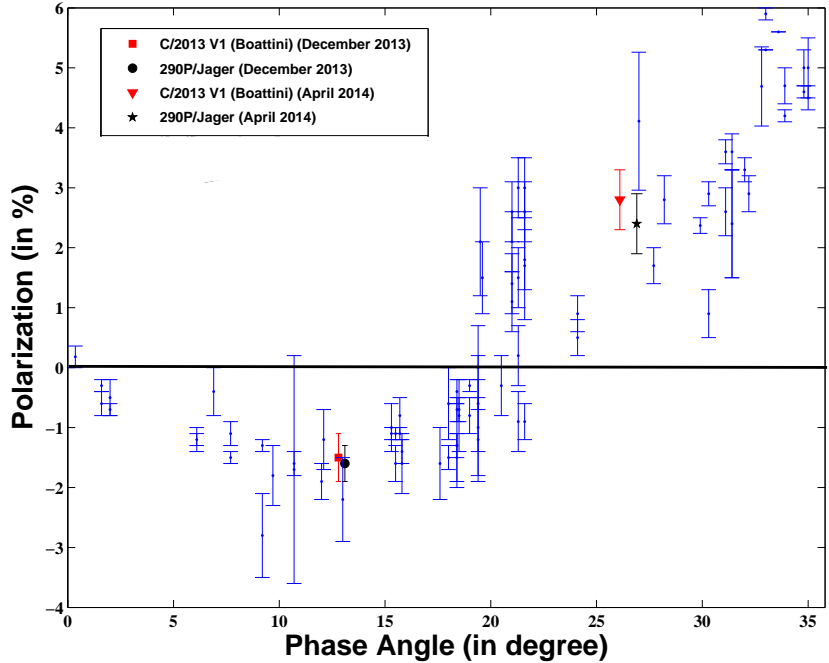}
\caption{Polarization versus phase angle plot for different observed comets at R filter. The linear polarization data of different comets are taken from the Database of Comet Polarimetry by Kiselev et al. (2006).}
\end{figure*}


\section{Discussion}
Cometary dust grains  evolve  with  the nucleo-centric distance and are not bound to be the same in the regular extended coma as in other specific cometary  feature like jets, fans etc. The CCD imaging polarimetry is an important tool to reveal the existence of different grains population in the cometary coma and also explores the physical evolutions of the dust grains in the inner and outer coma of the comet. This tool derives the realistic  characteristics  of the dust population of any specific  comet. The variation in the brightness and the polarization distribution over the entire coma indicates change in the physical properties of dust grains in different regions of the coma.

 The intensity of a comet is found to be higher in the near nucleus region due to high dust concentration. The azimuthally integrated brightness profile of comets C/2013 V1 (Boattini) and 290P/Jager follows a standard $(-1.0\pm0.08)$ slope variation with the increase of projected diameter suggests the proper subtraction of background sky counts which is a typical task in the study of comets. The fall in the brightness profile in sunward and anti-sunward directions is being observed for both the comets in all observational run which suggests the variation of physical properties of dust grains in due time. The decrease in intensity is slightly higher in the antisolar direction for comet C/2013 V1 (Boattini) at small photocentric distance on 2013 December 4 whereas a higher fall in intensity is being noticed on December 5 in the solar direction at an intermediate distance. Comet 290P/Jager shows higher fall in intensity in the solar direction during both the nights of December. In the month of April, the steep fall in the intensity is being noticed in the antisolar direction for both the comets. This fall is observed mainly in the outer coma. The solar radiation pressure changes the velocity outflow of the dust particles depending on their size pronounced in the outer coma of the comet which is the quite obvious reason for the steep fall in the intensity (Tozzi et al. 2004). The variation in the brightness profile in the considered direction is mainly due to the variation in the relative dust concentration at different regions of the cometary coma since the brightness strongly depend on the dust concentration in an optically  thin coma of the comet. Modulation of dust production rate, sublimation of high albedo icy grains due to the solar radiation pressure and variation in the dust physical properties might also cause the variation (Farnham et al 2009).

To enhance the special structure in the comet we treated the intensity images of both the comets C/2013 V1 (Boattini) and 290P/Jager using Larson-Sekanina's  rotational gradient  technique (Larson and Sekanina 1984) but no such special feature is being detected during both the periods. At lower phase angles the ejections mainly takes place in front of the observer in the direction of sun (Hadamcik \& Levasseur 2003). So it is very difficult to detect such feature in the comet with the spatial resolution of AST provides, as the jets usually have a reduced surface on the images.

The integrated aperture polarization value is found to be almost uniform with the increase of projected diameter for both the comets during the two phase of observations.  It has been observed that comet C/2013 V1 (Boattini) and Comet 290P/Jager show  high negative polarization of  -3 and -4.5 per cent respectively in the near nucleus region as compared to the whole coma polarization of  ($-1.4 \pm 0.3$)  and ($-1.6 \pm 0.3$) per cent in 2013 December at a phase angle of about 13$^\circ$. Ebisawa \& Dollfus (1993) proposed that the presence of submicronic size particles can explain this abnormal high negative polarization. The positive polarization observed in the outer coma of the comets 290P/Jager and C/2013 V1 (Boattini) in December is due to the presence of Rayleigh particles characterized by small scattering cross-section at optical wavelengths. It is to be noted that the dust grains of small sizes are expected to be accelerated at higher expansion speed as compared to the large compact particles due to their strong dynamical coupling to the gas outflow.

\section{Summary}
The polarimetric observations of comets C/2013 V1 (Boattini) and 290P/Jager were carried out with the AST during two periods: $(i)$ 2013 December 4 and 5 before perihelion passage and $(ii)$  2014 April 24 just after their  perihelion passage. The intensity profile of comet C/2013 V1 (Boattini) and 290P/Jager shows the asymmetric nature of the cometary coma. The variation of slopes along sunward and antisunward direction explores the various physical evolutions going in the cometary coma. The deviation of the brightness profile in the solar and antisolar direction from the standard canonical nature confirms the time variable dust outflow which is mainly due to the modulation of dust production rate, sublimation of high albedo icy grains due to the solar radiation pressure and variation in the dust physical properties.

The degree of polarization of comets C/2013 V1 (Boattini) and 290P/Jager is found to be $-1.4$$\pm 0.3$ and $-1.6$$\pm 0.3$ per cent at about 13$^\circ$  phase angle;  and $+2.8$$\pm 0.5$\% and $+2.5$$\pm 0.5$\% at 27$^\circ$  phase angle. The polarization maps showed a good agreement with the computed polarization value. It has been noticed from polarization maps that the polarization is highly negative in the circumnucleus halo of comet 290P/Jager as compared to that of comet C/2013 V1 (Boattini)  during December observation while a high positive polarization is being observed at the near nucleus region of both the comets in April observation.

\section{Acknowledgement}

Through this acknowledgment, we express our sincere gratitude to ARIES, Nainital, India for allocation of observation time. The reviewer of this paper Professor Alan Fitzsimmons is highly acknowledged for constructive comments which definitely helped to improve the quality of the paper. This work is supported by the Science and Engineering Research Board (SERB), a statutory body under Department of Science and Technology (DST), Government of India, under Fast Track scheme for Young Scientist (SR/FTP/PS-092/2011). PDR wants to acknowledge DST INSPIRE scheme for the fellowship.

\end{document}